\documentclass[trackchanges]{aastex701}

\usepackage{graphicx}
\usepackage{amssymb}
\usepackage{epstopdf}
\newcommand{\kms}{\,km\,s$^{-1}$}
\usepackage{newtxtext,newtxmath}
\usepackage[T1]{fontenc}

\DeclareRobustCommand{\VAN}[3]{#2}
\let\VANthebibliography\thebibliography
\def\thebibliography{\DeclareRobustCommand{\VAN}[3]{##3}\VANthebibliography}

\usepackage{amsmath}	% Advanced maths commands
\usepackage[export]{adjustbox}
\usepackage{xcolor}

\shorttitle{CO(3-2) Observations of Haro 2}
\begin{document}

\title{High Resolution Observations of CO(3-2) in Haro 2: Cool Molecular Outflows of a Ly$\alpha$ Emitter}
\correspondingauthor[show]{Sara Beck}
\author[orcid=0000-0002-5770-8494, gname=Sara, sname=Beck]{Sara Beck}
\affiliation{School of Physics and Astronomy, Tel Aviv University\\
Ramat Aviv, Israel 69978}
\email[show]{becksarac@gmail.com}

\author[orcid=0000-0001-9155-3978, gname=Pei-Ying, sname=Hsieh]{Pei-Ying Hsieh}
\affiliation{National Astronomical Observatory of Japan; Academica Sinica Insitute of Astronomy and Astrophysics, Taipei 106216, Taiwan} 
\email{pei-ying.hsieh@nao.ac.jp}

\author[orcid=0000-0003-4625-2951, gname=Jean, sname=Turner]{Jean Turner}
\affiliation{Department of Astronomy, UCLA, Los Angeles, Ca. USA } 
\email{turner@astro.ucla.edu}

\begin{abstract}
Haro 2 is a blue compact dwarf galaxy and the closest (at 21Mpc)  known $Ly\alpha$ emitter. UV and optical observations have found Haro 2 to be immersed in shells of partly ionized gas expanding at $\sim200$\kms~.  Observations of CO(2-1) (\citet{Beck2020}; Paper 1) with moderate ($2^{\prime\prime}$) resolution discovered a large-scale one-sided outflow associated with a soft X-ray bubble and apparently driven by the young star clusters created in the recent starburst.  We present here SMA observations of CO(3-2) with $1.1^{\prime\prime}$ resolution in Haro 2. The opposite-side component of the large molecular outflow is detected lying in the direction of the fast ionized outflow; it is very confined in area.   An  additional outflow is apparent in the CO(3-2); it is south-east of the galaxy in a region holding weak star formation and a moderate-luminosity hard X-ray source.    All the molecular outflow and filament velocities are within $\lesssim\pm50$\kms~ of the systemic velocity of the galaxy. 
\end{abstract}


\section{Introduction}
Haro 2 (Mkn 33, Arp 233, UGC 5720; D$\approx 21 Mpc, v_\odot 1440$\kms) is one of the most luminous blue compact dwarf galaxies in the Local Universe.   Its optical and UV spectrum has strong emission lines and Wolf-Rayet features (\citet{Mendez00}, \citet{Lequeux95},\citet{Kunth98}), and the infrared spectrum \citep{Matsuoka2012}, and strong compact radio continuum emission \citep{Beck00} indicate recent and on-going star formation concentrated in bright star clusters or groups of clusters along the galactic major axis. 
      
Haro 2 is immersed in, or has created, remarkable gas structures.  The Lyman~$\alpha$ line has the P Cygni profile  of an outflow, with absorption and emission features offset $\pm\sim200$\kms~( \citet{Lequeux95},\citet{Kunth98}).The red-shifted $Ly\alpha$ emission has been localized to the west of the galaxy \citep{Mas03}, and is probably associated with the   filaments and shells of $H\alpha$ seen in that region by \citet{Mendez00} and in HST Legacy Archive images.  The X-ray observations of \citet{Summers01} and \citet{OtiFloranes12} find a complex halo of emission, probably a hot super-bubble or bubbles, extending both east and west of the galaxy.   Observations of molecular and atomic gas in Haro 2  by \citet{Bravo03} and \citet{Meier01} at low spatial resolution found warm ($T_k>20~K$) molecular gas and a peculiar velocity field, with the main kinematic and photometric axes apparently misaligned. SMA observations of CO(2-1) with $\sim2^{\prime\prime}$ resolution, reported in Paper 1,  showed that part of the unusual  velocity field arises in a one-sided molecular outflow north-east of the galaxy, coincident with the hot super-bubble observed by \citet{Summers01} and probably driven by one of the young giant star clusters.  

There are many questions concerning the gaseous envelope of Haro 2.  Is the molecular gas around the hot super-bubble cold material that was entrained with the hot gas, or is it plasma that has cooled down?  What is the current state of the outflow gas?  Is there another 'horn' of outflow on the other side of the galaxy, and are the molecular outflows related at all to the ionized gas seen in $Ly\alpha$ and $H\alpha$? 
How has this dwarf galaxy created a network of extra-galactic gas greater than that of many more luminous starbursts in larger galaxies, and what will be the possible effects on further star formation and the evolution of the galaxy?  

These issues motivate the observations we report here of the CO(3-2) line in Haro 2.   With the higher spatial resolution (about twice that of Paper 1) and the chance to probe a warmer component of molecular gas, we obtain a fuller picture of the state of the gas component in this remarkable galaxy.  
\section{Mapping the Data }
Haro 2 was observed on 19 January 2021 with the Sub-Millimeter Array in the Extended configuration and on 20 March 2021 in the Compact configuration. Combining data from these two arrays gives final beams with major axis from 0.7$^{\prime\prime}$ to 1.2 $^{\prime\prime}$, depending on the weighting used. Vesta and 0958+655 were respectively the flux and phase calibrators, 3C84, 3C279 the bandpass calibrators, and dual receivers (rx345, rx400 Ghz) were used.  The data were binned by 4 channels to a final velocity channel of 4.95 Mhz or 4.2\kms~ at CO(3-2), calibrated and reduced with MIR\_IDL in the usual fashion.  The data were then cleaned and mapped in CASA. 

The observations covered a range from 342.5 to 344.5 Ghz, including the frequency CO(3-2), which was detected, and CS, HCS, HCN, HCO+, which were not.  The $870\mu$m continuum was searched for but not detected at greater than $\sim2\sigma$ and will not be discussed further here.  

The CO(3-2) maps displayed here were mapped with the auto multi-threshold mask and 500-1000 niter. This route was chosen to minimize some negative bowl structures appearing in the data.  The (3-2) maps discussed are natural weighted and have 1.1$\times$1.1$^{\prime\prime}$ beams, almost a factor of two higher resolution than the 1.9$\times$1.65 $^{\prime\prime}$ beams of the CO(2-1) observations. 
\begin{figure}
    \centering
    \includegraphics[scale=0.5]{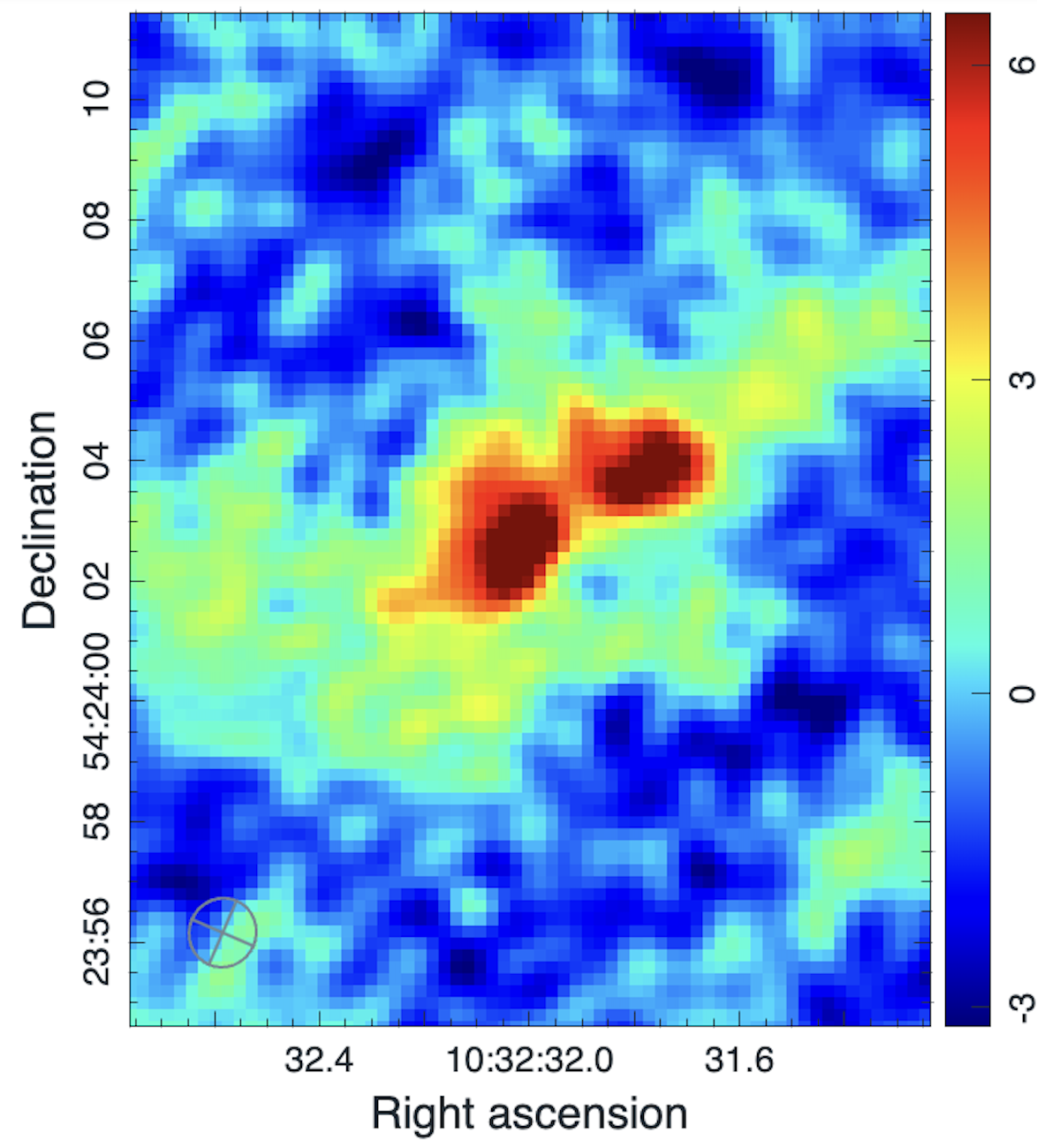}\\

    \includegraphics[scale=0.6]{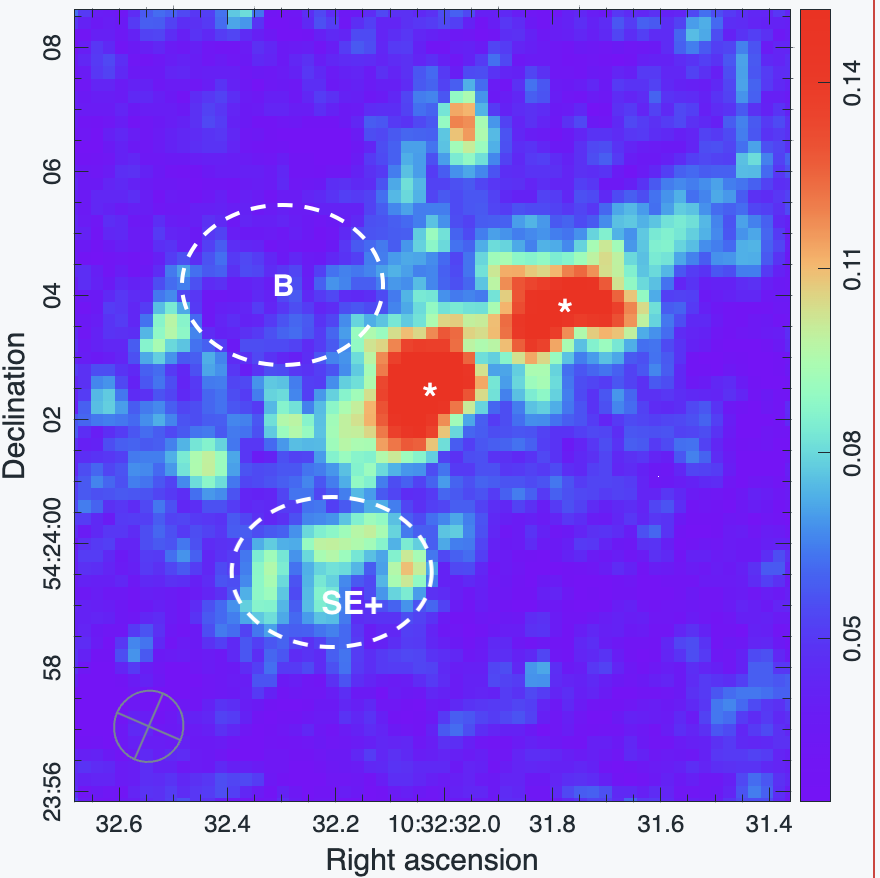}
    \caption{Top: The integrated intensity (moment 0)  of the CO(3-2) line on Haro 2. The color bar has units $(Jy/bm)(km/s)$. The beam is shown in the lower left. Bottom: The peak intensity (moment 8) of CO(3-2) on Haro 2.  Color bar units are $(Jy/bm)$. The two great starburst clusters are marked with small white stars, the superbubble or outflow region of hot gas is indicated with a white oval and letter 'B' and the south-east extended star formation region with a white oval and  'SE+'} 

    \label{fig:enter-label}
\end{figure}

\subsection{CO(3-2) Distribution}
The top panel of Figure 1 shows the total integrated CO(3-2) line intensity over the full frequency range in a stretch emphasizing weak extended emission; the lower panel is the moment 8 map of peak intensity, which emphasizes clumps. 

The two great starburst regions (marked in the figure) dominate the CO(3-2) emission.  These giant star clusters or groups  of star clusters, each of $L_{tot}\sim10^9 L_\odot$, are discussed in Paper 1 and in \citet{Beck00}.  The CO(3-2) line profiles and structure of the northern source indicate that it probably includes multiple sub-sources that cannot be isolated with this resolution and the southern source appears uniform.  The extended CO(3-2) emission north-west of the galaxy is associated with the expanding superbubble discovered in Paper 1 and so marked.  

A significant feature of the CO(3-2) map which was not apparent in the CO(2-1) results is the  clumpy emission south-east of the main body of the galaxy.  This SE CO(3-2) emission region is quite significant; the brightest clump is $\sim$ half the intensity of the southern star cluster source. This region is discussed in section 6.  

\begin{figure}
\begin{center}
\includegraphics[width=1\linewidth]{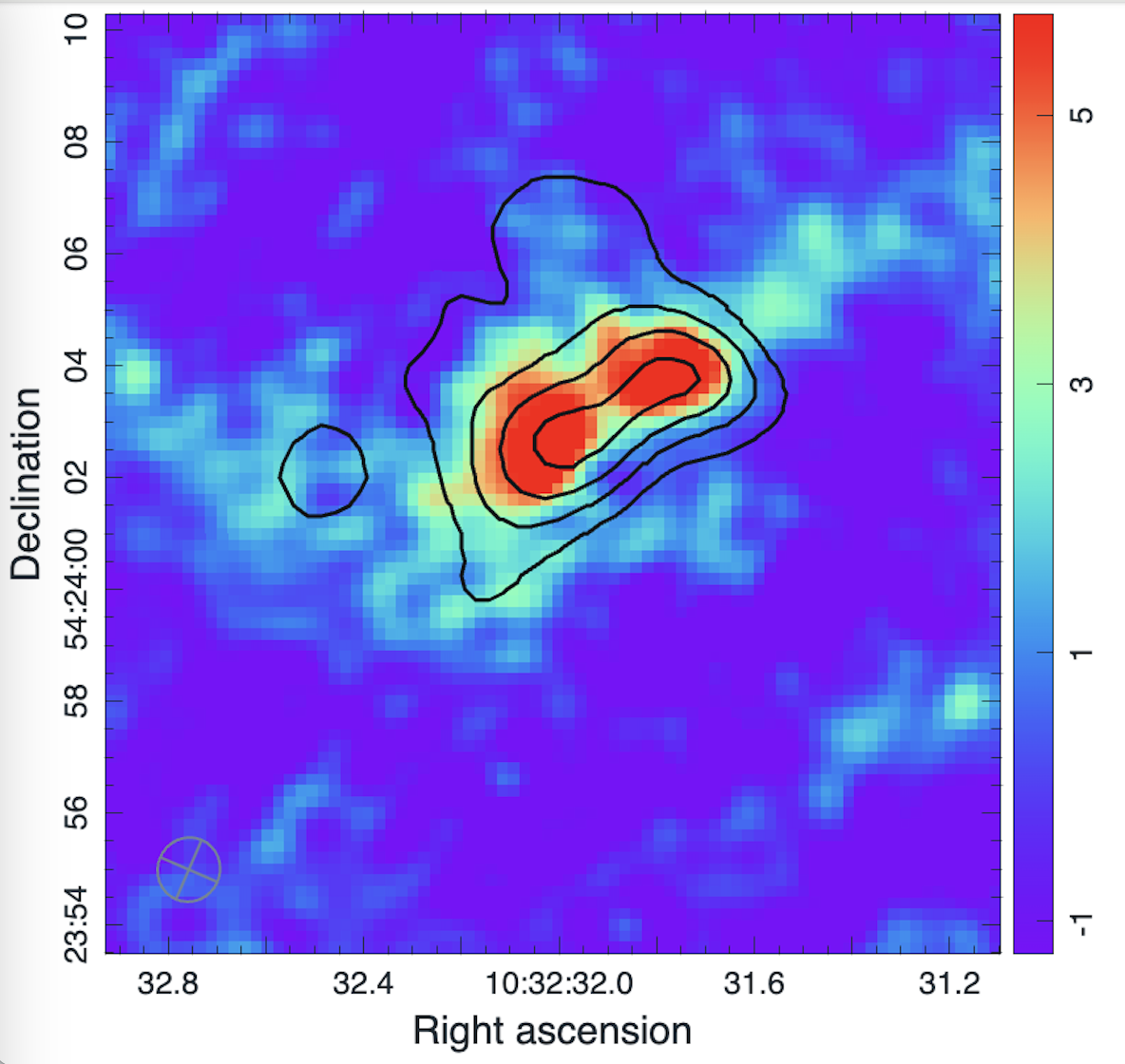}
  
\caption{Moment 0 of the CO(3-2) map in color, with color bar units $(Jy/bm)(km/s)$, and moment 0 of the CO(2-1) emission in contours at levels $4,6,8,10 (Jy/bm)(km/s)$}.
  \label{fig:enter-label}
\end{center}
\end{figure}

How does the CO(3-2) distribution compare to the CO(2-1) in Paper 1?  Figure 2 overlays the two images, with the CO(3-2) at native resolution.   The two maps agree well over the two bright clumps that form the main body of the galaxy. The relative strengths of CO(2-1)and CO(3-2), and their implication for the state of the molecular gas, are discussed in Section 5.
 \section{CO(3-2) Kinematics: Outflows and Filaments} 
 The gas, both hot and cold, in and around Haro 2 has multiple complex velocity structures. The CO(3-2) observations can measure the kinematics of the molecular gas with higher spatial resolution than earlier work, and are better probes of the somewhat warm molecular gas often associated with intense star formation.  
 
 In Figure 3 we display the first moment (weighted mean velocity) of the CO(3-2) in Haro 2.  In the main body of the galaxy, the two great starburst clusters are offset in velocity, with the stronger south-east cluster $\sim40$\kms~ red of the northern cluster; this agrees with Paper 1 and the H$\alpha$ results of \citet{Mendez00}. 
 North-east of the galaxy, the 'superbubble' region appears as 'horns' offset in velocity. Opposite to the superbubble, on the south-west of the galaxy,  we see a smaller but definite pair of velocity horns. Finally, a complex velocity gradient in the SE region at the southern tip of the galaxy marks what we believe to be another outflow cone.  
 
 These three regions of molecular gas outside the main body of the galaxy are marked in the first panel of Figure 4 with dotted lines on the velocity gradients. Position-Velocity Diagrams (PVDs) were obtained along these lines and  are displayed in the other panels.  

 We will examine the superbubble and its counterpart in section 4.1  and the SE region  section in 4.2.

 \subsection{The North-East Outflow and its Far-Side Counterpart } 
 The 'superbubble' or 'North-East outflow' is the largest molecular outflow in Haro 2. It appears in Figure 3  as two 'horns' at distinct velocities north and east of the 
 main body of the galaxy, and in the PVD of Figure 4 with a structure typical of expanding shells or cones. In Paper 1 it is shown that this north-eastern molecular outflow is associated with a bubble of hot X-ray emitting gas. It is suggested there that the bubble or outflow is driven by or starts from the southern of the starburst clusters in the main body of the galaxy.    
 
 The first moment (Figure 3) shows that the weak extensions opposite to the 'superbubble' on the south and west of the galaxy (seen in the moment 0 and 8 maps of Figure 1 around RA 10:32:31.6 RA, Dec54:24:02) correlate with distinct  velocity features.  We believe this emission to be the other (according to the redshift, the far) side of the outflow whose near side is the great 'superbubble', and note that it also present in the CO(2-1) data.
 
 This suggested counterpart to the NE outflow is significantly weaker and less spatially extended than is the opposite side.  In contrast, the  ionized gas outflows are strong and extended in the SW and are not seen NE. \citet{Mas03} model the ionized flows of Haro 2 as resulting from a shell of neutral gas expanding at $\sim200~km~s^{-1}$ into a low-density ionized halo.  The CO data suggest that the SW region is mostly filled with ionized gas and the molecular component is limited to very close to the galaxy. 
 
 \begin{figure}
\includegraphics[width=1\columnwidth]{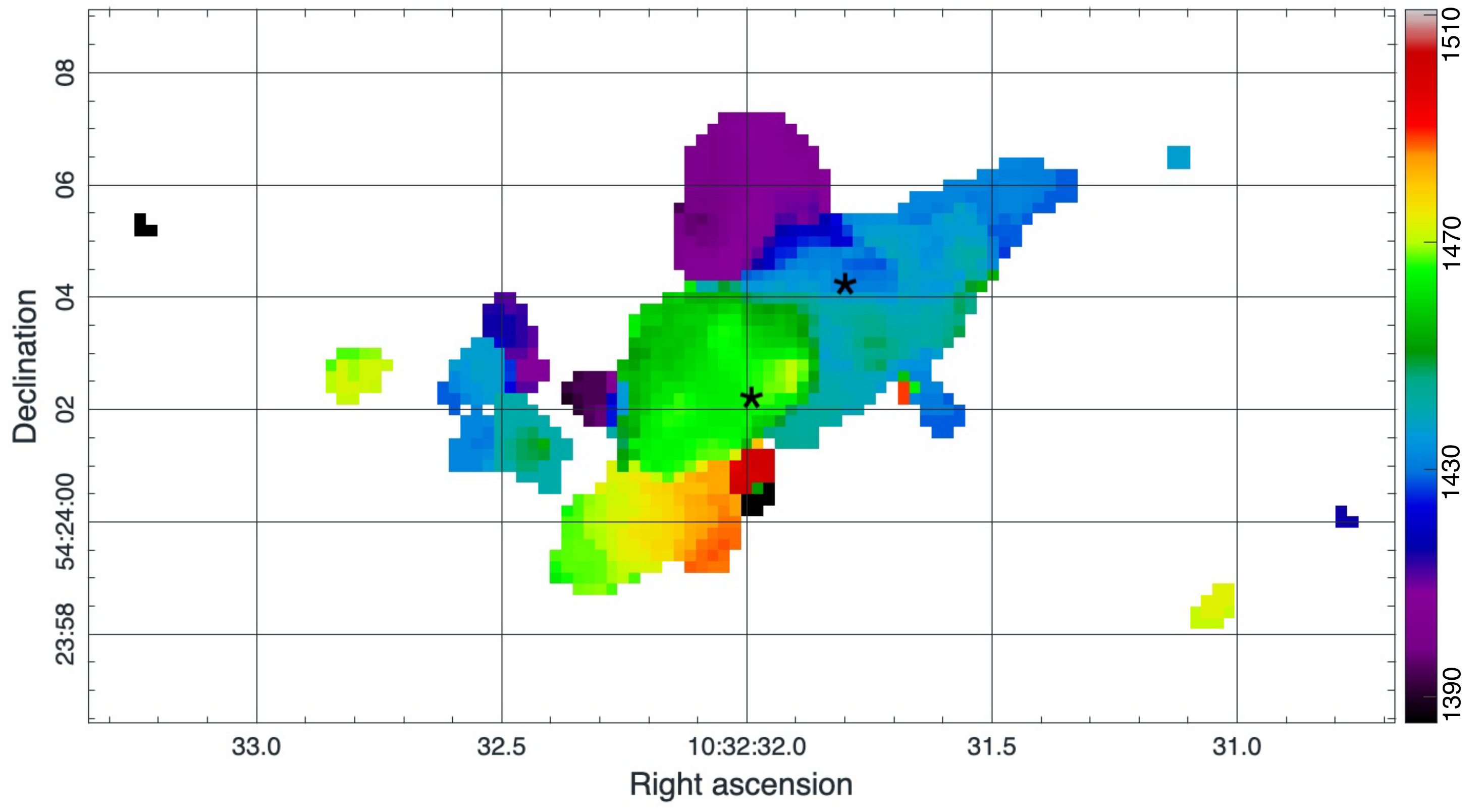}
\caption{First moment or weighted mean velocity of CO(3-2) in Haro 2.  Units of the color bar are velocities in (Km/s).   The data cube was clipped at $3.5\times10^{-2}Jy/bm$. The great star clusters are marked with black stars. The outer border of the mapped region corresponds roughly to the $0.8(Jy/bm)(Km/s)$ level of the zero moment map.   }
\end{figure}

\begin{figure}

      \centering
      
          \includegraphics[width=1\linewidth]{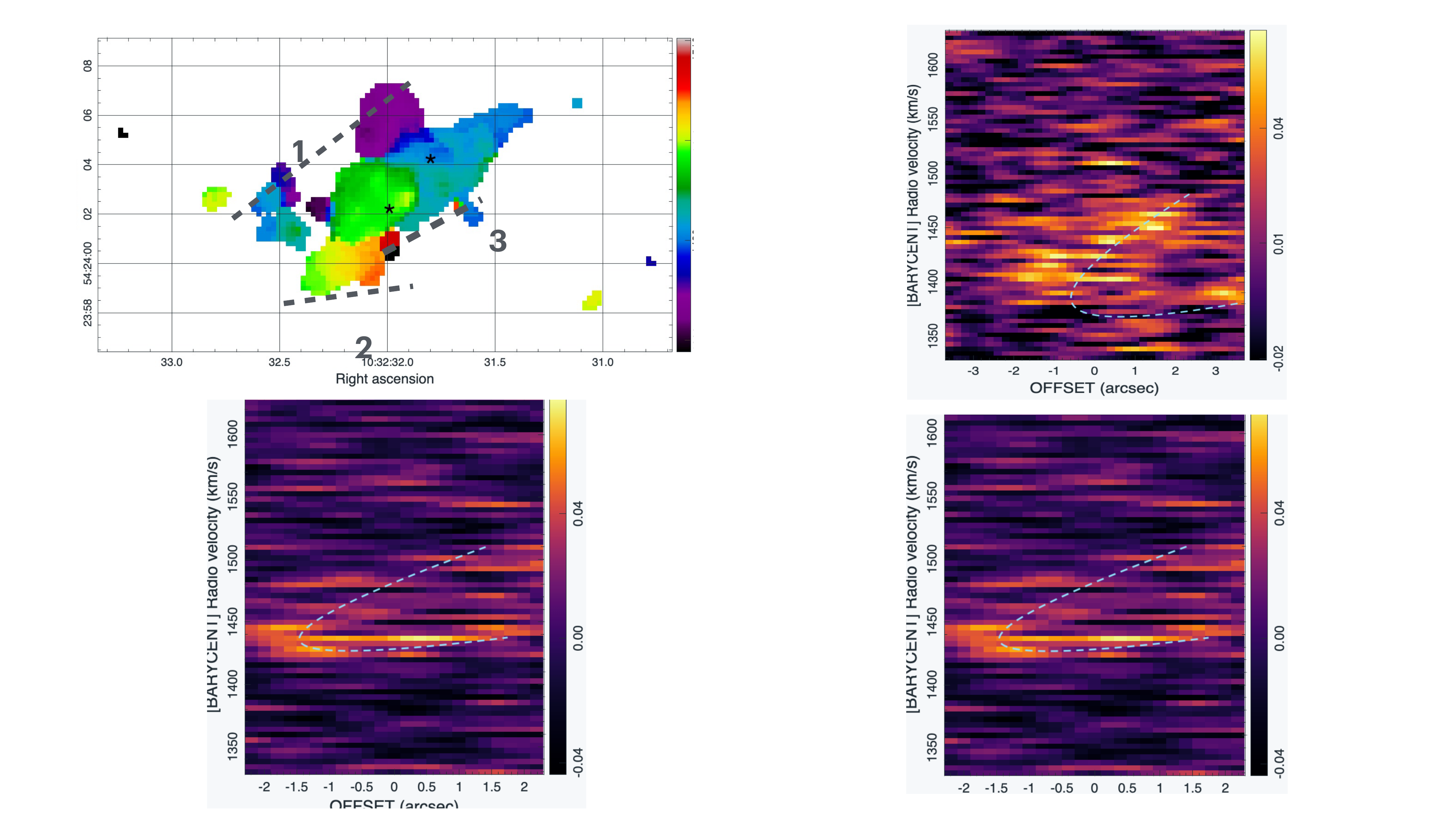}
          \caption{Top Left: The first moment map of CO(3-2), as in the previous figure, marked with three dotted lines showing where  Position-Velocity Diagrams were obtained. The positions are identified with numbers 1 for the NE 'superbubble', 2 for the SE region, and 3 for the SW structure that we believe to be the other side of the 'superbubble'. Clockwise from top left: the Position-Velocity Diagram on each marked line. Units of the color bar are $(Jy/bm)$ and units of Offset are distance in arcseconds from the center of the line.  Dashed curves are sketched on each PVD to accentuate the velocity structure.   }
          \label{fig:enter-label}
      \end{figure}

\subsection{A New Outflow Cone on the South-East}
The SE region has spatially distinct velocity features suggestive of an outflow over a $\sim45$km/s range west-east.  The centroid of this outflow is at $v_\odot\sim1480$\kms~. This is offset to the red with respect to the great star clusters, consistent with the velocity gradient north-south across the galaxy.   The moment 1 map of Figure 3 and the PVD in Figure 4 shows an expansion or outflow signature and a velocity gradient from west to east across this region. 
This velocity gradient is also evident in the spectra of Figure 5, which show
double-peaked profiles shifting over 45\kms from west to east.

%\begin{figure}
%\includegraphics[width=1\columnwidth]{mkn33-2025/CO32.png} 
%\caption{Top left: The CO(3-2) emission in the 1475.65 \kms channel. Bottom Left: the peak intensity (moment 8). Top right: the velocity (moment 1).  In all three maps the line shows where the P-V diagram, displayed on the bottom right, is calculated, and the circles where the spectral profiles of Figure 5 were recorded.    }
%\end{figure}

\begin{figure}
    \centering
    \includegraphics[width=1\linewidth]{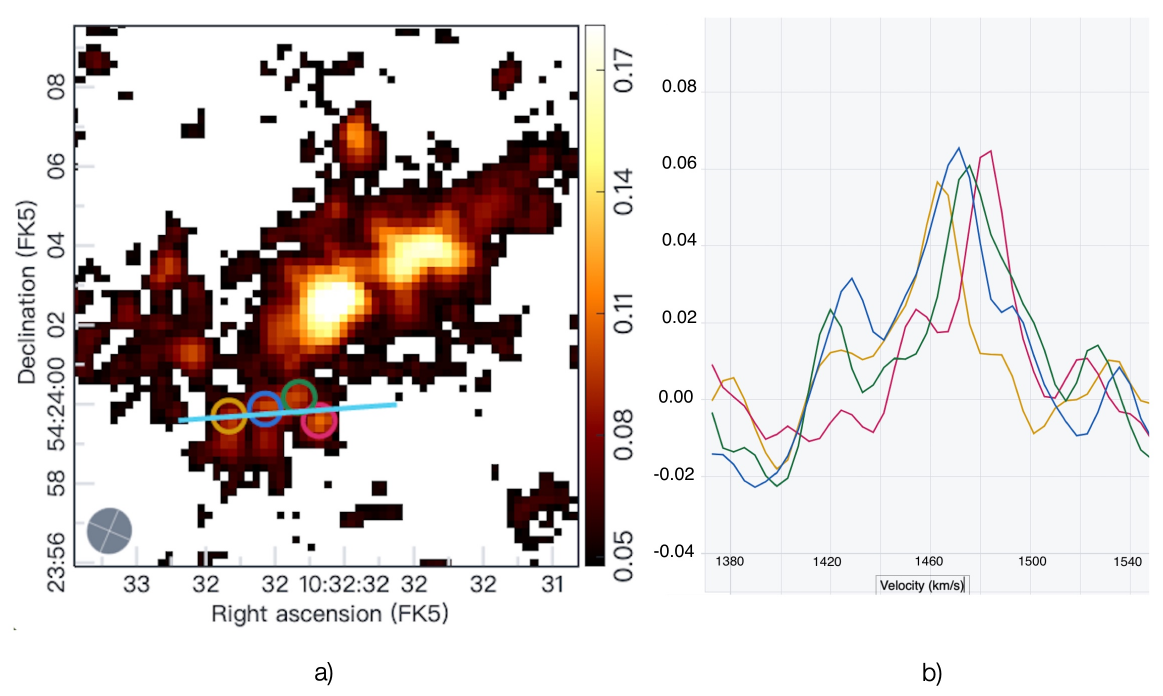}
                \caption{Left: The moment 8 map with the positions of the spectral profiles marked with beam-sized circles.  The colors of the circles correspond to the colors of the spectral profiles on the right. Right: The spectral profiles of the CO(3-2) line in the positions given on the right. The X axis is velocity in \kms~ and the Y axis intensity in Jy/bm. 
    The order is yellow-blue-green-red east to west.  }
    \label{fig:enter-label}
\end{figure}
We see that the SE region is the site of a molecular outflow or expanding shell with $\sim45$\kms~ velocity spread.  Why is this feature not apparent in the CO(2-1) data? One reason is that the SE cone is small, $\sim3^{\prime\prime}$ from edge to edge, and could not be well distinguished with the $\sim2^{\prime\prime}$ spatial resolution of the CO(2-1) maps. Another factor is that the CO(3-2) is enhanced relative to the CO(2-1) here (as may be seen in Figure 2). If the $R_{32}$ ratio (discussed in Section 5) reflects the molecular gas temperature,  the SE gas is warmer than that entrained around the 'superbubble' outflow and therefore stronger in CO(3-2)

Unlike the main 'superbubble' outflow NE of the galaxy, the SE molecular outflow  does not correlate with extended hot gas.  \citet{OtiFloranes12} found a hard X-ray source which they call X2 in the middle of the SE region. This source has total counts $8.9\times10^{-4}s^{-1}$,  a factor  $\sim8$ fainter than the source on the bright SE cluster that \citet{OtiFloranes12} label X1, and is weaker than the  $10^{39}erg/s$ of ultra-luminous X-ray sources.  The X2 X-ray source is probably an Intermediate (IMXB)  X-ray binary; these need massive young donor stars and are typically found in  star formation regions.  X2 has not formed a X-ray bubble, and there is at present no evidence connecting X2 and the SE molecular outflow except their spatial proximity. 
\section{Gas Conditions}
The total flux of the galaxy at CO(3-2) is $45\pm2$ Jy(km/s) in a $5^{\prime\prime}$ radius; the uncertainty  reflects possible influence of bowl structure in the maps.  Of this 32-35 Jy(km/s) (depending on the brightness level chosen as boundary) are in the bright clumps of the galactic plane. Fluxes in the extended emission regions are   5-7 Jy(km/s) in the superbubble north-east of the galactic plane and 7-10 Jy(km/s) in the SE emission region; these values are totals in polygon regions and the ranges are our best estimate of how well the main body of the galaxy can be excluded. 

Comparing to the CO(2-1) maps of Paper 1, we see that  CO(3-2) is concentrated on the embedded star clusters: the bright clumps make up $\sim3/4$ of the total flux at (3-2) but only $\sim1/3$  of the CO(2-1) total.  The total CO(2-1) is dominated by the outflows, and the CO(3-2) emission by the brightest young star clusters.  

The ratio $R_{32}=I(3-2)/I(2-1)$ is $1\pm0.15$ for the bright clumps of the plane. This ratio is typical of that in compact star formation regions \citep{Meier01} where the gas is heated by the strong stellar radiation. $R_{32}$ is  $0.5\pm0.1$ for the SE region;  somewhat lower than on the bright clumps,suggesting that the gas is cooler on the SE region, which has weak star formation,  than in the very bright main clusters.  Both the bright clumps and the SE region have warmer gas than does the main outflow, discussed in the next section. 
 \subsection{The State of Molecular Gas in the Main Outflow} 
  What is the state of the molecular gas on the main outflow, above the galactic plane? This outflow is coincident with a bubble of hot gas whose temperature derived from X-ray measurements is $1.77\pm 0.19\times10^7 K$ \citep{Summers01}.   We saw above that the galaxy's total CO(3-2) emission is dominated by the young star clusters, and the CO(2-1) emission by the bubble region;  $R_{32}$ on the bubble is $0.23\pm0.03$, significantly lower than in the star formation regions of the galactic plane or in the smaller outflow in the SE region. The simplest interpretation is that the gas is cooler on the superbubble than in the star formation regions (although the optical depth of the CO lines is not known and it is possible that the excitation of one or both lines may be sub-thermal).   Note that the relative distribution of the CO(2-1) and CO(3-2) and the concentration of CO(3-2) on the star clusters is similar to that seen in the nucleus and multiphase outflow of the much more luminous starburst NGC 253 \citep{Krieger19}. 
 
Molecular gas outflows are common in galaxies, galaxy clusters, and individual star formation regions (\citet{Veilleux20},\citet{Russell19},\citet{Leaman19},\citet{Bekki10} ). The outflows are often 'mutiphase', including warm and hot gas with the cool molecular material,  as we see in Haro 2. % bubble is clearly associated with the young star clusters in the plane, and the 
 The hot and warm gas in the Haro 2 bubble outflows is presumably heated by the SNe, stellar winds or other energetic activity in the young clusters.   What is the source of the molecular gas?   \citet{Veilleux20}) set out three possible mechanisms (which can co-exist in one object) for the creation and maintenance of the molecular component in multiphase outflows: 1) the molecules form in situ as the hot gas cools and recombines, 2) the molecules are entrained and carried along by the hot fluid 3) the molecules are directly accelerated by radiation or cosmic rays.  
 
While the molecules in the Haro 2 outflow do appear concentrated on the edges of the bubble, where the hot gas is cooling down, it is unlikely that they formed in situ.  First, because hot gas will not recombine into molecules until it has cooled and there is no evidence, such as e.g. $H\alpha$, for warm gas in the bubble (the ionized gas shells reported by \citet{Mendez00} are 
{\it{west}} of the galaxy). Second, because molecular recombination is a slow process \citep{Veilleux20}and the activity in Haro 2 is young.   The upper limit for the travel time of the outflow, which assumes that its velocity has been constant with no deceleration, is less than a few $10^7$ years, and the deep obscuration of the star clusters which drive the outflow argue that they are similarly young.   The direct acceleration model is also unlikely to apply in Haro 2 as the young star clusters are too weak and too highly obscured to provide sufficient radiation pressure.   

The most probable mechanism at work is that the cluster winds have entrained some of the molecular gas near the embedded star clusters and have lifted it out of the galaxy plane.  Even though the molecular gas on the clusters was relatively warm and strong in CO(3-2), it will cool before going very far:  \citet{Silich23} showed that dense cluster winds will cool from some hundreds of $K$ to $\sim20~K$ in a few pc travel (while the clusters in that paper are brighter and more concentrated than those in Haro 2, the cooling time should be on the same order).  
 
\section{The SE Emission Region: Weak Star Formation, an Outflow and an X-ray Source}
The CO(3-2) observations found  unexpected levels of spatial complexity and high-velocity kinematics, including a molecular outflow or bubble, in the SE emission region.  Other features of this part of Haro 2:  

 \begin{itemize}
\item The radio continuum measured at 22 Ghz \citep{Ferraro25} with a $0.95^{\prime\prime}$ beam  and at 5 Ghz with a $0.59\times0.47^{\prime\prime}$ beam {\footnote{5 Ghz data from the  historical VLA program AY35}} shows 'scruff': no distinct statistically significant sources but extended weak emission. The total emission is  $0.6mJy$, about $7\sigma$,  summed over $1^{\prime\prime}$
 diameter at 5 Ghz and $0.3 mJy$ at 22 Ghz. For comparision, the  bright star clusters have 3.6, 2.8 mJy in the same area.   
\item Near-infrared images at $1.6\mu$m and $2.2\mu$m and the WFPC2 image at $H\alpha$ from the HST Legacy Archive show extended scruffy emission of the same overall shape as the CO(3-2), with  
bright clumps that are probably small star clusters, and total emission about 1/5 that of the main SE cluster. 
\item $5.8\mu$m and $8.0\mu$m images from Spitzer IRAC show scruffy emission similiar to the radio continuum. 
\end{itemize} 

All the tracers agree with a picture of SE as a region of moderate star formation, with an extended low-level background of emission and a few star clusters, smaller and weaker than those in the main body of the galaxy. 
There is no obvious candidate in SE for the source driving the small outflow.  
   
 \section{Summary and Discussion}  
 We present maps with $1.1^{\prime\prime}$ spatial resolution
and 4.3 km/s spectral resolution of CO(3-2) emission in the dwarf starburst Lyman alpha emitter
Haro 2. We compare the images to the known distribution
of H$\alpha$, Ly$\alpha$,X-rays, radio continuum, and CO(2-1), and find the following:
\begin{itemize}
    \item We detect CO(3-2) emission over a region of >$15^{\prime\prime}$(1.5 kpc), 
 centered on the $L_{tot}\sim10^9 L_\odot$ starforming clusters. 
3/4 of the total 33$\pm$2 Jy km/s CO(3-2) flux
 is within 200 pc of the star clusters; the rest in extended emission around the galaxy. 

 \item The CO(3-2)/CO(2-1) ratio $R_{32}$ has the typical starburst value on the star clusters and is lower, indicating cooler gas, in the extended emission. 

 \item The kinematics  of the CO(3-2) line is very complex. There is a blue to red gradient  from north to south on the star clusters, and three velocity structures, described below,  off the main body of the galaxy.   

\item A blue molecular outflow is seen to the northeast of the galaxy, on the
side opposite to the observed Lyman alpha outflow. This feature is associated with hot X-ray emitting gas.  It was also observed in CO(2-1) \citep{Beck2020}. 

\item The CO(3-2) maps show a new outflow centered $3-4^{\prime\prime}$  (300-400 pc)
southeast of the starburst clusters in a region of weak and diffuse star formation.   This source would have been difficult to distinguish in the lower resolution CO(2-1) maps. The total velocity range of the outflow is 45\kms~.  The source driving this outflow is not apparent. 

\item We do not see an extended outflow in CO(3-2) coincident with the Ly$\alpha$ and H$\alpha$ outflows to the southwest of
the starburst clusters. There are short filaments or horns of CO(3-2) emerging in that direction; their velocities and placement suggest that they are the opposite side of the large northeast outflow. 

The CO distribution suggests that molecular gas has mostly been cleared, perhaps by an older outflow,  from the western side of the galaxy.  Ly$\alpha$ emission is now seen from the cleared region. 
 \end{itemize}
 
 \subsection{Discussion: A Lucky View of a Moderate Starburst? } 
The high resolution CO(3-2) maps have refined the picture of molecular structures near and on Haro 2. We see filaments and outflows extending in all directions from the galaxy. The puzzle of Haro 2 is how a relatively modest dwarf starburst can host so much outflow activity.  We saw in Section 5 that there is at least as much molecular gas in the outflows ($\sim10^8M_\odot$ from observations of the dominant CO(2-1) level,\citet{Beck00})  as in the giant starforming clouds.  Haro 2 has total IR luminosity only $\sim7\times10^9L_\odot$~\citep{Beck00} but its outflows have molecular mass comparable to or greater than that in much brighter starbusts such as NGC 253 \citep{Bollato13}).  

Haro 2 became a well-studied target  because it is one of the very closest $Ly\alpha$ emitters. How does the molecular activity compare to the ionized? 

The total molecular mass in the outflows is $\sim10^8M_\odot$, approximately a factor of 10 greater than the ionized mass quoted by \citet{Lequeux95}; a ratio consistent with the rough relation found by \citet{Fluetsch19} for a sample of 45 Local Universe outflows. The ionized outflow seen in $Ly\alpha$ and $H\alpha$ is faster than the molecular activity and significantly more extended. The $H\alpha$ shells in the Hubble image reach $\sim$15 $^{\prime\prime}$ from the galaxy; the molecular outflows less than 5 $^{\prime\prime}$.  The total width to zero intensity of CO(3-2) is $\sim150$\kms~,  significantly less than the $\sim200$\kms~ expansion velocity of the $Ly\alpha$ from this galaxy; it is possible that the ionized and molecular flows are coeval and that the molecular gas has smaller extent because it is slower. 
 
 It is probable that we observe the Haro 2 starburst at a particularly propitious {\it time} and from a propitious {\it angle}. {\it Time} because gas the starburst expelled $\sim5-8\times10^6$ years ago (deduced from the Ly$\alpha$ velocity and extent) has travelled far enough to now be clear of the galaxy and can be observed; {\it angle} because the galactic plane is tilted so that the outflows are apparent.  Further observations, especially of the surrounding neutral gas into which these outflows are expanding, could be very valuable.   Haro 2 is a excellent laboratory for detailed study of galactic gas flows.

\end{document}